\documentclass[fleqn,usenatbib]{mnras}

\usepackage{newtxtext,newtxmath}

\usepackage[T1]{fontenc}

\DeclareRobustCommand{\VAN}[3]{#2}
\let\VANthebibliography\thebibliography
\def\thebibliography{\DeclareRobustCommand{\VAN}[3]{##3}\VANthebibliography}

\usepackage{graphicx}	
\usepackage{amsmath}	
\usepackage{url}
\usepackage[dvipsnames]{xcolor}
\usepackage{float}
\usepackage{physics}
\usepackage{booktabs}
\usepackage{tabularx}
\usepackage{multirow}

\title[FRB Spectro-Temporal Survey]{Validating the Sub-Burst Slope Law: A Comprehensive Multi-Source Spectro-Temporal Analysis of Repeating Fast Radio Bursts}

\author[K. Brown et al.]{Katie Brown,$^1$\thanks{E-mail: kbrow327@uwo.ca}
Mohammed A. Chamma,$^2$ 
Fereshteh Rajabi,$^2$
Aishwarya Kumar,$^1$
\newauthor Hosein Rajabi,$^3$ and Martin Houde$^1$\thanks{E-mail: mhoude2@uwo.ca}
\\
$^{1}$Department of Physics and Astronomy, The University of Western Ontario, 1151 Richmond Street, London, Ontario N6A 3K7, Canada\\
$^2$Department of Physics and Astronomy, McMaster University, 1280 Main Street West, Hamilton, Ontario, L8S 4L8, Canada\\
$^3$Department of Computer Science, The University of Western Ontario, 1151 Richmond Street, London, Ontario N6A 3K7, Canada
}

\date{Accepted 2024 January 29. Received 2024 January 13; in original form 2023 August 22}

\pubyear{2024}

\begin{document}
\label{firstpage}
\pagerange{\pageref{firstpage}--\pageref{lastpage}}
\maketitle

\begin{abstract}
We conduct a comprehensive spectro-temporal analysis of repeating Fast Radio Bursts (FRBs) utilizing nine distinct sources, the largest sample to date. Our data set includes 175 sub-bursts and 31 multi-component bursts from 11 data sets, with centre frequencies ranging from 149--7144 MHz and durations spanning from 73 $\mu$s--13 ms. Our findings are consistent with the predictions of the Triggered Relativistic Dynamical Model (TRDM) of FRB emission. We affirm the predicted quadratic relationship between sub-burst slope and central frequency, as well as a linear dependence of the sub-burst bandwidth on central frequency that is consistent with mildly-relativistic Doppler broadening of narrow-band emission. Most importantly, we confirm the sub-burst slope law, a predicted inverse relationship between sub-burst slope and duration, to hold consistently across different sources. Remarkably, we also discover that the drift rates of multi-component bursts follow the same law as the sub-burst slopes, an unexplained result that warrants further investigation. These findings not only support the TRDM as a viable framework for explaining several aspects of FRB emission, but also provide new insights into the complex spectro-temporal properties of FRBs.

\end{abstract}

\begin{keywords}
transients: fast radio bursts -- methods: data analysis -- relativistic processes -- radiation: dynamics -- radiation mechanisms: non-thermal
\end{keywords}



\section{Introduction}\label{sec:intro}

Fast Radio Bursts (FRBs) remain one of the most intriguing phenomena in astrophysics, captivating researchers since their discovery in 2007 \citep{Lorimer2007}. Characterized by their extremely short durations (on the order of a millisecond) and high energies (up to $\sim 10^{32}~\mathrm{erg\,s^{-1}\,Hz^{-1}}$ in isotropic luminosity), FRBs exhibit remarkable variability in their observed properties, further complicating their enigmatic nature \citep{Petroff2022}. Despite this diversity, it is widely assumed that these transient events originate from highly energetic processes, likely tied to compact objects \citep{lyutikov2016}. Although several theories, including magnetar flares and compact-object mergers, are considered promising, the underlying mechanism remains a mystery. 

One promising approach to gaining insights into the emission mechanism and source dynamics is through studying the spectro-temporal properties of FRBs. These include the spectral bandwidth, temporal duration and central frequency of a burst, as well as frequency drift rates between and within sub-bursts\footnote{When multiple resolved components are visible within a single waterfall or dynamic spectrum, these individual pulses are called 'sub-bursts'.}. The ‘sad trombone’ effect refers to the downward drift in frequency of consecutive sub-bursts (which we will refer to as the drift rate; \citealp{hessels2019,josephy2019}). In contrast, the sub-burst slope refers to the (typically steeper) downward frequency drift within a single sub-burst. 

Several relationships between spectro-temporal properties have been derived within the framework of the triggered relativistic dynamical model (TRDM), a simple model in which FRB emission is triggered by a signal from a background source, such as a magnetar or pulsar \citep{Rajabi2020, houde2019}. The TRDM was inspired by, but is not limited to, the application of superradiance to FRB emission \citep{Houde2018, houde2019}. The central prediction of the TRDM is the `sub-burst slope law,' an inverse relationship between the sub-burst slope and duration. The model also proposes several relationships tied to the central frequency: an inverse dependence on duration and a quadratic correlation with sub-burst slope, while we expect a further relationship with the bandwidth from the Doppler effect. These predicted patterns have been observed in FRB 20121102A, the primary focus of most studies due to its wealth of observational data \citep{Rajabi2020, Jahns2022, Chamma2022}. However, investigations exploring these relationships across multiple FRB sources have been few and constrained, typically limited by sparse data and narrow duration and frequency ranges \citep{Chamma2021,wang2022_freq}.

In this paper, we seek to bridge this gap by conducting a comprehensive spectro-temporal analysis of repeating FRBs utilizing nine distinct sources—the largest sample to date. By evaluating relationships among the slope, duration, frequency, and bandwidth of each burst and sub-burst, we aim to test the predictions of the TRDM within a broad and diverse data set. This investigation, which encompasses a frequency range akin to the entire detected FRB population, enables a robust assessment of these hypothesized laws. Whether our study uncovers ubiquitous patterns across all sources and bursts or identifies significant deviations or exceptions, our findings will provide valuable insights into the FRB population and, ultimately, their origins. 

This paper is organized as follows: in Section \ref{sec:methods} we outline our analysis approach, including burst waterfall preparation and dispersion measure (DM) considerations, and discuss the data used. Our findings and their implications are presented in Section \ref{sec:results}, and the key insights are summarized in Section \ref{sec:summary}.

\section{Methods} \label{sec:methods}

\subsection{Analysis}\label{sec:analysis}

Turning to the details of our approach, the first step involves preparing the burst waterfalls for measurements. This includes performing down-sampling in frequency and/or time, background subtraction, and masking of noisy channels and radio interference. Each step is handled on a case-by-case basis.  For processing and spectro-temporal measurements, we use the graphical user interface \textsc{Frbgui}\footnote{\texttt{https://github.com/mef51/frbgui}}, developed by \cite{Chamma2022}. 

Next, we consider the dispersion measure (DM) for each FRB source, an element that can strongly affect the measured sub-burst slope. For this reason, we dedisperse each burst over a range of trial DMs, following the methodology of \cite{Chamma2021}. We adopt a range of $20$ pc cm$^{-3}$, centered on the reported DM value. To minimize the change in sub-burst morphology between consecutive DMs, we use a conservative step size of $0.2$ pc cm$^{-3}$.

To increase the Signal-to-Noise ratio (S/N), reduce the impact of burst structure and interference, and allow for more consistent measurements, we perform a 2D autocorrelation function analysis on each waterfall \citep{hessels2019,Hewitt2023}. Each autocorrelation is fit with an ellipsoid with a 2D Gaussian intensity distribution, from which we extract measurements of sub-burst slope, duration, and bandwidth; this process is detailed in \cite{Chamma2022}. 

When analyzing bursts with multiple components, vertical (i.e. temporal) divisions are used to separate them into their sub-bursts. Each sub-burst then undergoes the same analysis as an individual burst. The complete burst is also characterized through the autocorrelation process, with the measured slope being the drift rate between consecutive components. 

The first step in analyzing the measurement results is determining both the range of possible DMs and the representative DM value for each source. We begin with the 20 pc cm$^{-3}$ range in DM for each source and restrict it by eliminating any DM that is found to be invalid for any of the bursts. A DM is considered too high if it produces a positive sub-burst slope or a fitting error larger than 40\% (usually occurring where the bursts are nearly vertical; \citealt{Chamma2022, Jahns2022}). Within the TRDM framework, positive slopes (which result in bursts that appears to be upward-sloping) are unphysical and produced by over-dedispersion. Conversely, we consider bursts as under-dedispersed if they become curved or so stretched that fitting an ellipsoid to their autocorrelation is unfeasible.

As explained in \citet{Chamma2022}, the limited range of DMs is used to calculate the representative DM for each source, which is taken to be that which yields the best fit for the sub-burst slope law predicted by the TRDM of \cite{Rajabi2020}. To determine this, a fit of the form $|\dd\nu_{\mathrm{obs}}/\dd t_{\mathrm{D}}|\nu_{\mathrm{obs}}^{-1}=A/t_{\mathrm{w}}$, with $\nu_\mathrm{obs}$ the sub-burst's central frequency, $t_{\mathrm{w}}$ its duration and $t_{\mathrm{D}}$ the delay time before its arrival (see equation 1 below), is found for the measurements taken at each DM within the predetermined limited range. For each source, we select the DM that results in the fit with the reduced-$\chi^2$ closest to unity as the representative DM. We then plot the measurements at these representative DMs to study the relationships between the spectro-temporal quantities. The `uncertainty' of the sub-burst slope in the measurements presented in Section \ref{sec:results} is set by the corresponding values obtained over the extent of the limited DM range about the representative DM.

\subsection{Data}\label{sec:data}

\begin{table*}
\centering
\caption{Summary of data analyzed and key results of this study. The number of bursts used for most sources is less than the total detected due to the exclusion criteria of low S/N and severe truncation. The sub-burst slope law coefficient $A$ (see equation \ref{eq:slopelaw}) is determined for each source. The DM range in the sixth column is the restricted set of values that produces reasonable dedispersion; see Sec.~\ref{sec:analysis}.}\label{tab:sources}
\begin{tabular}{llllllll}
\toprule
\textbf{FRB Source} & \textbf{Telescope} & \textbf{\begin{tabular}[c]{@{}l@{}}Frequency Range\\ (MHz)\end{tabular}} & \textbf{\begin{tabular}[c]{@{}l@{}}Number of \\ Bursts Used\end{tabular}} & \textbf{\begin{tabular}[c]{@{}l@{}}Sub-Burst \\ Measurements\end{tabular}}& \textbf{\begin{tabular}[c]{@{}l@{}}DM Range\\ (pc cm$^{\vb{-3}}$)\end{tabular}} & \textbf{\begin{tabular}[c]{@{}l@{}}Representative \\ DM (pc cm$^{\vb{-3}}$)\end{tabular}}   & \textbf{Coefficient \textit{A}} \\ \midrule
20121102A           & Arecibo            & 1150--1730                                                               & 18                                                                        & 24                                                                         & 558.0-559.6 & 558.0                                                                               & 0.076 $\pm$ 0.004                \\
                    & Arecibo            & 4100--4900                                                               & 16                                                                        & 19 & 556.0-560.0                                                                        &                                                                                     &                               \\
                    & Green Bank         & 4000--8000                                                               & 17                                                                        & 22 & 555.0-561.0                                                                        &                                                                                     &                               \\
20180916B           & uGMRT              & 550--750                                                                 & 12                                                                        & 16 & 348.2-348.6                                                                        & 348.6                                                                               & 0.11 $\pm$ 0.01                 \\
                    & LOFAR              & 110--188                                                                 & 12                                                                        & 12 & 348.4-348.8                                                                        &                                                                                     &                               \\
                    & CHIME              & 400--800                                                                 & 13                                                                        & 21 & 348.4-348.8                                                                        &                                                                                     &                               \\
20201124A           & Effelsberg         & 1200--1520                                                               & 12                                                                        & 20 & 410.6-411.2                                                                        & 410.6                                                                               & 0.17 $\pm$ 0.02                 \\
20180301A           & FAST               & 1000--1500                                                               & 7                                                                         & 12 & 514.4-516.0                                                                        & 515.4                                                                               & 0.088 $\pm$ 0.007               \\
20180814A           & CHIME              & 400--800                                                                 & 4                                                                         & 7 & 186.0-186.8                                                                         & 186.8                                                                               & 0.09 $\pm$ 0.2                  \\
20200929C           & CHIME              & 400--800                                                                 & 4                                                                         & 6 & 411.8-412.8                                                                         & 412.4                                                                               & 0.08 $\pm$ 0.2                  \\
20190804E           & CHIME              & 400--800                                                                 & 4                                                                         & 5 & 361.0-363.2                                                                         & 363.2                                                                               & 0.11 $\pm$ 0.03                 \\
20190915D           & CHIME              & 400--800                                                                 & 5                                                                         & 6 & 487.6-488.8                                                                         & 487.6                                                                               & 0.12 $\pm$ 0.02                 \\
20201130A           & CHIME              & 400--800                                                                 & 5                                                                         & 5 & 287.2-287.8                                                                         & 287.8                                                                               & 0.29 $\pm$ 0.07                 \\ 
\bottomrule
\end{tabular}
\end{table*}

In total, our analysis covers nine repeating FRB sources, encompassing 11 data sets collected from six different telescopes. Our measurements are obtained from 128 bursts, among which 31 were divided into multiple sub-bursts, resulting in a data set of 175 single pulses. The analyzed data have centre frequencies ranging from 149--7144 MHz, and their durations span 73 $\mu$s--13.22 ms. Information on each source and data set is summarized in Table \ref{tab:sources}.

For the data sets where not all detected bursts were utilized, exclusions were made due to insufficient S/N for reliable autocorrelation analysis, or in instances where parts of the bursts were severely truncated in frequency with their structure extending beyond the observed bandwidth. No correction is applied to account for scattering.

\subsubsection{FRB 20121102A}
FRB~20121102A is one of only two known periodic repeaters, with a period of approximately 160 days \citep{Rajwade2020,cruces2021}. The source has been localized to a star-forming region in a dwarf galaxy at $z=0.1927$, and is associated with a persistent radio source \citep{Tendulkar2017}. We use three sets of data from this source. 

\cite{hewitt2021} observed 478 bursts during a burst storm in 2016 with the Arecibo telescope over a band of 1150--1730 MHz (these data will be referred to as `20121102A Arecibo-1'). We select a subset of 18 bursts from this data set to avoid this source skewing the overall results. We choose waterfalls with clearly visible signals and no severe truncation, preferentially selecting bursts with multiple components. We divide four of them into sub-bursts, leading to a total of 24 single pulses.

The second data set, `20121102A Arecibo-2,' comes from \cite{michilli2018}, who observed this source using Arecibo over 4.1--4.9 GHz. All 16 bursts from this data set are used, with three of them divided into sub-bursts, yielding 19 measurements.

Lastly, we use 17 of the 21 bursts detected by \cite{gajjar2018} with the Green Bank telescope over 4--8 GHz, the highest frequency detection of FRB~20121102A to date. Three of these bursts are split into two to four sub-bursts yielding 22 single pulses. For the data from \cite{michilli2018} and \cite{gajjar2018}, we use the results from the processing and measurements of \cite{Chamma2022} who follow an identical procedure, except for the DM grid.

After analysis of these three data sets together, the representative DM value for FRB~20121102A is established at 558.0 pc cm$^{-3}$, consistent with \cite{Chamma2022}.

\subsubsection{FRB 20180916B}

FRB 20180916B, another periodic repeater with a period of about 16 days \citep{Amiri2020}, resides in a star-forming region within a nearby (149 Mpc) massive spiral galaxy \citep{marcote2020}. We use three data sets for this source.

\cite{marthi2020} observed 15 bursts with the upgraded Giant Meterwave Radio Telescope (uGMRT) over 550--750 MHz. We use 12 bursts, dividing four into multiple components, yielding measurements for 16 individual sub-bursts.

The second data set is from \cite{pleunis2021}, who detected 18 bursts from FRB~20180916B using the Low Frequency Array (LOFAR) over a band of 110--188 MHz; this is the lowest frequency detection of any FRB to date. We use 12 of these bursts in our analysis.

Lastly, we use 19 bursts from The Canadian Hydrogen Intensity Mapping Experiment (CHIME) over a band of 400--800 MHz \citep{Andersen2019}. We take measurements from 13 bursts, splitting four into sub-bursts for a total of 21 single pulses.

Collectively, these data sets lead us to determine the representative DM for FRB~20180916B as 348.6 pc cm$^{-3}$, consistent with previously reported values \citep{Nimmo2021,Chamma2021,Gopinath2024}.

\subsubsection{FRB 20201124A}

FRB 20201124A is the third closest repeating FRB source, located in a massive star-forming galaxy at $z=0.098$ \citep{Nimmo2022}. \cite{Hilmarsson2021} detected 20 bursts from FRB~20201124A with the Effelsberg 100-m telescope over 1200--1520 MHz. We use 12 of these bursts and divide four of them into three sub-bursts each, resulting in a final data set of 20 individual sub-bursts. Upon analysis, we establish the representative DM for this source as 410.6 pc cm$^{-3}$, close to the value of 411.6 pc cm$^{-3}$ reported in \citet{Hilmarsson2021}.

\subsubsection{FRB 20180301A}

FRB~20180301A, observed by \cite{luo2020} using The Five-hundred-meter Aperture Spherical radio Telescope (FAST) over 1000--1500 MHz, has been localized to the outskirts of an intermediate-mass star-forming galaxy at $z=0.33$ \citep{Bhandari2022}. From the 15 bursts, we selected the seven with an S/N sufficient for the autocorrelation process to produce a visible ellipse with a clear enough structure to be fit with a 2D Gaussian, and divided four of them into sub-bursts, resulting in 12 single pulses. We determined the representative DM for this source to be 515.4 pc cm$^{-3}$, which is consistent with the structure-maximization values reported in \citet{luo2020} (i.e., ranging from 515.9 to 518.3 pc cm$^{-3}$). The higher S/N-maximization value of 522 pc cm$^{-3}$ reported in \citet{Price2019} was found to produce over-dedispersion in numerous sub-bursts in our sample.

\subsubsection{Other FRB Sources From CHIME}

The remaining five sources were observed with CHIME over 400--800 MHz; these include FRB~20180814A \citep{Amiri2021}, and FRB~20200929C, FRB~20190804E, FRB~20190915D and FRB~20201130A \citep{Andersen2023}. We analyzed four bursts from both FRB~20180814A and FRB~20200929C, dividing two bursts from each into sub-bursts. From FRB~20190804E and FRB~20190915D, we analyzed four and five bursts, respectively, dividing one burst from each source into two sub-bursts. For FRB~20201130A, we used five bursts, none of which were split into sub-bursts.

Upon analysis, we established the representative DM values for these sources as follows: 186.8 pc cm$^{-3}$ for FRB~20180814A, 412.4 pc cm$^{-3}$ for FRB~20200929C, 363.2 pc cm$^{-3}$ for FRB~20190804E, 487.6 pc cm$^{-3}$ for FRB~20190915D, and 287.8 pc cm$^{-3}$ for FRB~20201130A.

\section{Results and Discussion}\label{sec:results}

\begin{figure*}
    \centering
    \includegraphics[width=\textwidth]{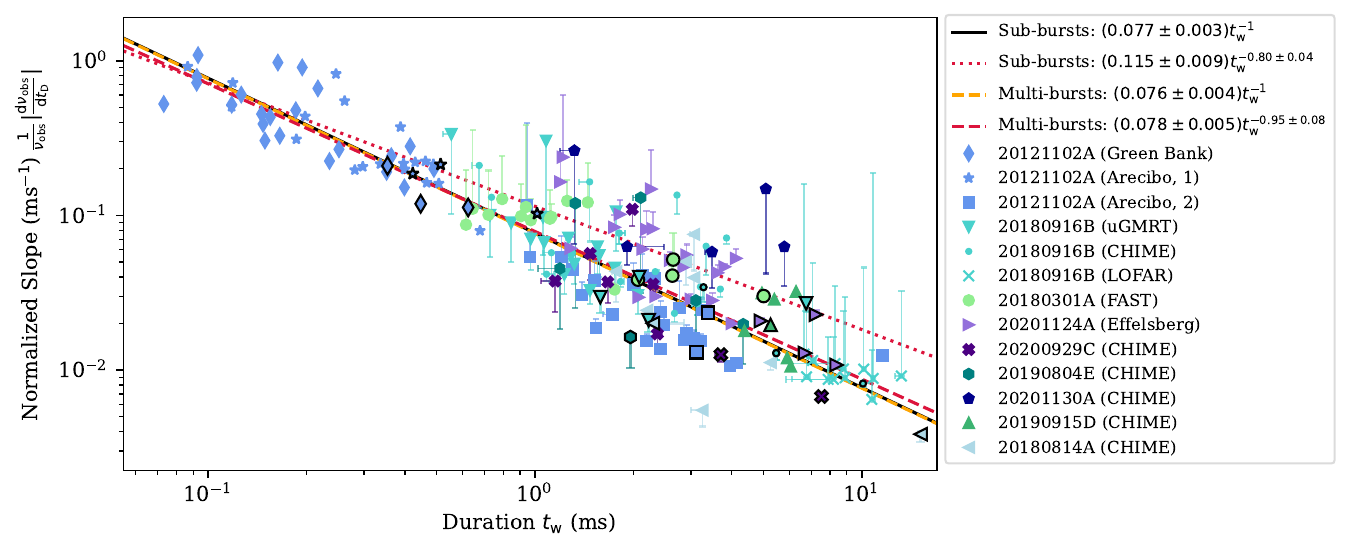}
    \caption{The sub-burst slope law and drift rate relation. Unoutlined points represent individual sub-bursts, whereas outlined points correspond to drift rate measurements for bursts with multiple components. The TRDM's inverse fit for sub-bursts is represented by the solid black line (partially hidden by the dashed orange line; see below), while the dotted red line shows a free power law fit. For multi-component bursts, the dashed orange and red lines display the inverse and power-law fits, respectively. In addition to the fit lines, the legend indicates the unique markers representing each data set (divided by FRB source and observatory; consistent between each figure). Each point is plotted at the representative DM determined for their source, with the error bars demonstrating the parameter's variation over the range of valid DMs; sources with limited DM ranges will have small or negligible error bars. We find that both sub-bursts and multi-component bursts closely follow the predicted inverse trend. }
    \label{fig:slopelaw}
\end{figure*}

\subsection{Sub-Burst Slope Law and Drift Rate Relation}\label{sec:slopelaw}

The key prediction of the TRDM, the sub-burst slope law, is an inverse relationship between frequency-normalized sub-burst slope and duration \citep{Rajabi2020}. This relationship is expressed as
\begin{equation}\label{eq:slopelaw}
    \frac{1}{\nu_{\mathrm{obs}}}\dv{\nu_{\mathrm{obs}}}{t_{\mathrm{D}}}=-\left(\frac{\tau_{\mathrm{w}}'}{\tau_{\mathrm{D}}'}\right) \frac{1}{t_{\mathrm{w}}}=-\frac{A}{t_{\mathrm{w}}},
\end{equation}
where, as noted before, $\nu_{\mathrm{obs}}$ represents the observed frequency, $t_{\mathrm{D}}$ the delay time, and $t_{\mathrm{w}}$ the observed duration. The coefficient $A\equiv\tau_{\mathrm{w}}'/\tau_{\mathrm{D}}'$ depends on the proper duration ($\tau_{\mathrm{w}}'$) and delay time ($\tau_{\mathrm{D}}'$), both governed by the properties of the source and emitting medium. We expect $A$ to be fairly constant for a given source, although there may be some spread between sources \citep{houde2019, Rajabi2020}.

The validity of the sub-burst slope law has been reinforced by several studies of repeating FRBs. \cite{Rajabi2020}, \cite{Jahns2022}, and \cite{Chamma2022} demonstrated its applicability to FRB 20121102A, analyzing 24, 849, and 167 bursts, respectively. Furthermore, \cite{Chamma2021} examined 42 bursts from FRBs 20121102A, 20180916B, and 20180814A, providing evidence that the law holds across varied sources. Preliminary evidence also suggests a similar law applies to the drift rates of multi-component bursts (see Figure 8 of \citealt{Jahns2022} and Figure 5 of \citealt{Chamma2022}), although definitive conclusions are hindered by a scarcity of measurements and narrow duration ranges.

As illustrated in Figure \ref{fig:slopelaw}, the findings for our nine sources align with these previous studies.  We present two fits: the predicted TRDM inverse relationship $At_{\mathrm{w}}^{-1}$, and a free power law of the form $A_{\text{free}}t_{\mathrm{w}}^{n_1}$. The inverse fit yields a coefficient $A=0.077 \pm 0.003$, consistent with $A=0.078 \pm 0.006$ reported for the multi-source study by \cite{Chamma2021} and $A=0.084$ obtained by \cite{Jahns2022} for FRB 20121102\footnote{\cite{Jahns2022} fit their sub-burst slope vs. duration data with the inverse formulation of the slope law, finding a coefficient of $b=-0.00862(37)$ MHz$^{-1}$. \cite{Chamma2022} convert this to a form equivalent to the form used here, finding $A=0.084$.}. Although \cite{Chamma2022} found a higher coefficient of $A=0.113 \pm 0.003$ for FRB 20121102A, they suggested potential skewing by high-slope outliers. We also find coefficients of $A_{\text{free}}=0.115 \pm 0.009$ and $n_1=-0.80 \pm 0.04$ for the power law fit, consistent with the predicted inverse trend. 

Individual fits for each of the seven sources with more than five measurements yields $A$ coefficients between $0.07-0.17$, with five centered closely around $0.1$ within a range of $\pm0.02$ ($A$ values for each source are included in Table \ref{tab:sources}). Despite slight deviations from the global trendline by certain sources, the overall data align well with the predicted inverse relationship. These results strongly reinforce the ubiquity of the sub-burst slope law in repeating FRBs, demonstrating its consistency across different sources.

Figure \ref{fig:slopelaw} also includes drift rate measurements from multi-component bursts, highlighting their consistency with individual sub-bursts. Notably, we found the drift rate measurements follow the same inverse trend as the sub-burst slopes, with $A_{\text{drift}}=0.076 \pm 0.004$ for the inverse fit, and $A_{\text{drift, free}}=0.078 \pm 0.005$ and $n_{\mathrm{drift}}=-0.95\pm0.08$ for the free fit. This relationship is not fully explained by the TRDM, depending on some unknown physical parameter (see Sec 2.1.3 in \citealp{Rajabi2020}). This closeness between $A$ and $A_{\text{drift}}$ is remarkable because it implies that the phenomenon responsible for the scaling within a sub burst also rules the scaling within a group of them. 

\begin{figure}
    \centering
    \includegraphics[width=\columnwidth]{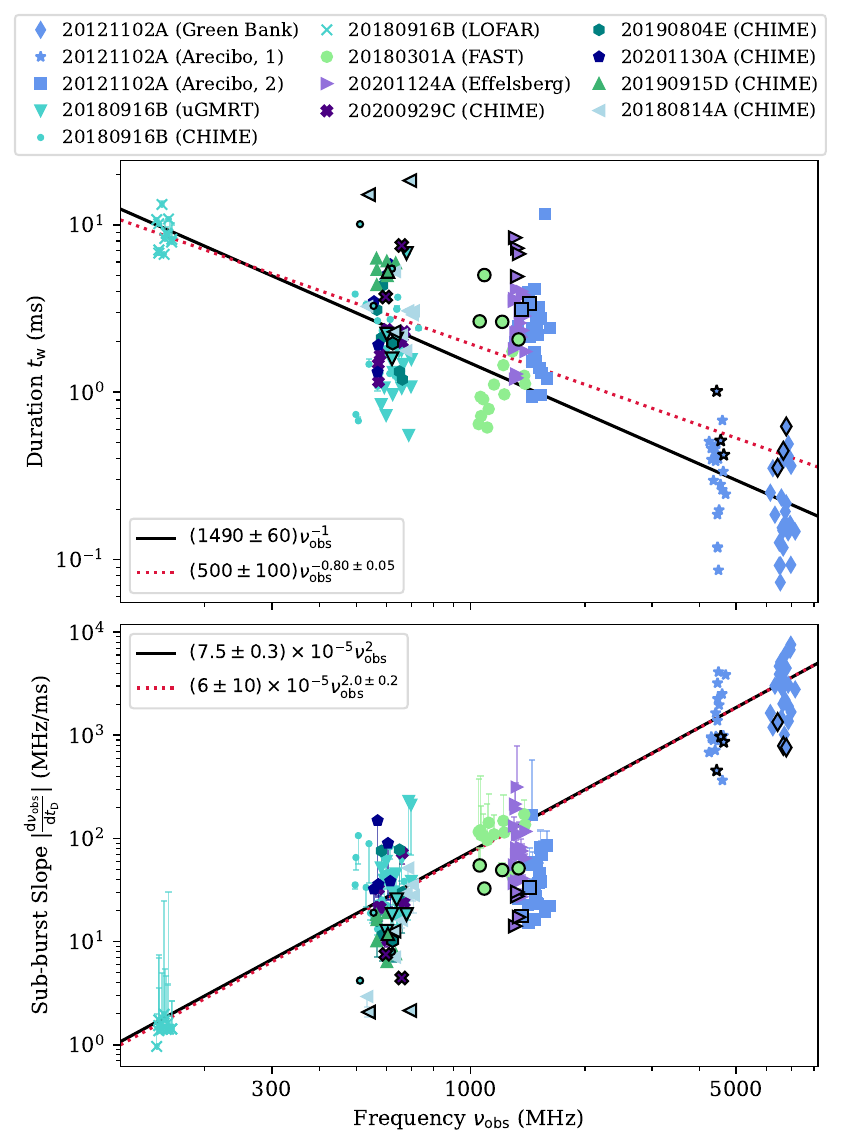}
    \caption{Relationships between the sub-burst duration (top) and the sub-burst slope (bottom) with frequency. The bursts show general trends of shorter durations and steeper slopes at higher frequencies. The black lines are the fits predicted by the TRDM, while the dotted red lines are free power law fits. Multi-component bursts are included for comparison (the outlined points; these are drift rate measurements rather than sub-burst slopes) but are excluded from the fits as we do not expect the same scaling between duration and frequency (and therefore also between the slope and frequency) for these bursts when compared to sub-bursts.}
    \label{fig:duration-freq}
\end{figure}

\subsection{Duration vs. Frequency and Slope vs. Frequency Relationships}

The TRDM also predicts an inverse relationship between sub-burst duration and frequency, articulated as 
\begin{equation}\label{eq:duration-freq}
   t_{\mathrm{w}}=\left(\tau_{\mathrm{w}}'\nu_0\right)\frac{1}{\nu_{\text{obs}} }=\frac{B}{\nu_{\text{obs}} },
\end{equation}
\citep{Rajabi2020}. In this equation, $\nu_0$ is the FRB-frame frequency of emission, assumed to be intrinsic to all FRB emission. Since the proper duration $\tau_{\mathrm{w}}'$ depends on the properties of the FRB environment, for a given $\nu_0$ we expect each source to potentially demonstrate different scaling for this relationship \citep{houde2019}. Furthermore, within each source we expect to see a significant spread in $\tau_\mathrm{w}'$, corresponding to a large vertical scatter in $t_\mathrm{w}$ vs. $\nu_\mathrm{obs}$. Previous studies \citep{gajjar2018, Chamma2022} found this inverse trend to hold for FRB 20121102A, but a multi-source study remains to be conducted.

\begin{figure}
    \centering
    \includegraphics[width=\columnwidth]{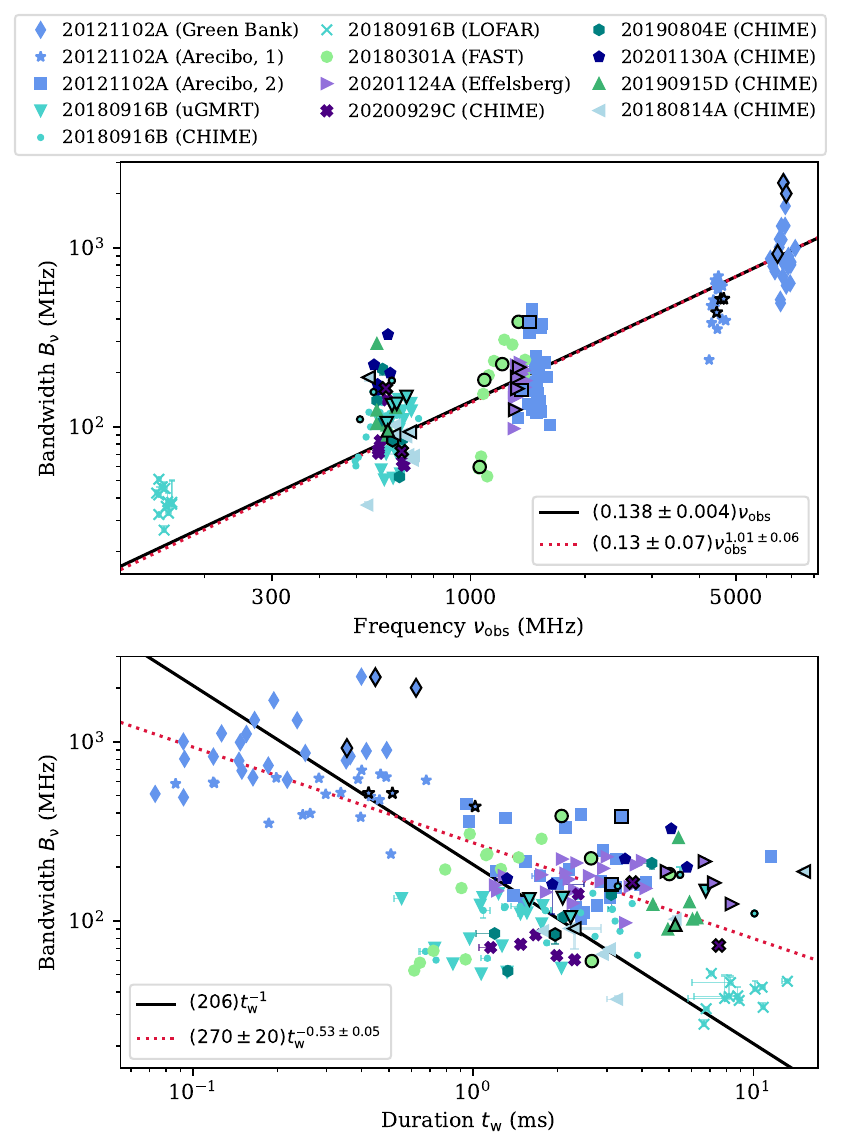}
    \caption{Sub-burst bandwidth relationships with frequency (top) and duration (bottom). In the upper panel, the solid black line represents the TRDM-predicted linear fit, while the dotted red line indicates the free power law fit. It shows a trend where bursts exhibit broader bandwidths at higher centre frequencies, consistent with mildly-relativistic Doppler broadening of narrow-band emission. In the lower panel, the solid black line is the inverse function derived from previously discussed relationships, and the dotted red line is a free power law fit. This demonstrates a dependence of bandwidth on duration approximating to the power of $-1/2$. Outlined points are multi-component bursts presented for comparison but not included in the fits.}
    \label{fig:doppler}
\end{figure}

Combining equations (\ref{eq:slopelaw}) and (\ref{eq:duration-freq}) yields a quadratic relationship between sub-burst slope and central frequency:
\begin{equation}\label{eq:slope-freq}
    \dv{\nu_{\text{obs}}}{t_{\mathrm{D}}}=-\frac{A\nu_{\text{obs}}^2}{\tau_{\mathrm{w}}'\nu_0}\equiv- C\nu_{\text{obs}}^2
\end{equation}
\citep{Chamma2021}. In this equation, we have introduced a new constant $C=1/(\tau_{\mathrm{D}}'\nu_0)=A/B$.  Similar to the inverse duration-frequency relationship, the source-specific properties influencing $\tau_{\mathrm{D}}'$ suggest that each source should potentially exhibit a unique $C$ value. We also anticipate significant scatter in  $\dd\nu_{\mathrm{obs}}/\dd t_{\mathrm{D}}$ due to the intrinsic spread in $\tau_\mathrm{D}'$. \cite{Chamma2022} found this law to hold for FRB 20121102A, and an overall trend between multiple sources was observed by \cite{Chamma2021} and \cite{wang2022_freq}.

These relationships – the dependencies of sub-burst duration and slope on frequency – are illustrated in Figure \ref{fig:duration-freq}.  We model the duration-frequency data using the TRDM's predicted inverse relationship in equation (\ref{eq:duration-freq}). This yields a coefficient of $B=1490 \pm 60$ ms MHz, closely matching the $1474 \pm 48$ ms MHz reported for FRB 20121102A by \cite{Chamma2022}. A power law fit $B_{\text{free}}\nu_{\text{obs}}^{n_2}$ yields $B_{\text{free}}=500 \pm 100$ and $n_2=-0.80 \pm 0.05$, consistent with the predicted inverse trend. 

We then fit the sub-burst slope vs. frequency data with the predicted quadratic trend $C\nu_{\text{obs}}^{2}$, finding $C=(7.5 \pm 0.3)\times10^{-5}$ ms$^{-1}$ MHz$^{-1}$. The free power law fit $C_{\text{free}}\nu_{\text{obs}}^{n_3}$ agreed with the quadratic trend, with $C_{\text{free}}=(6\pm10)\times10^{-5}$ and $n_3=2.0 \pm 0.2$. Both results are similar to the coefficient of $(6.3 \pm 0.3)\times10^{-5}$ ms$^{-1}$ MHz$^{-1}$ found by \cite{Chamma2022}. Significantly, these coefficients closely resemble the predicted value of $C=A/B=5.2\times10^{-5}$ ms$^{-1}$ MHz$^{-1}$, thereby affirming the consistency among the relationships established between sub-burst slope, duration and frequency (equations \ref{eq:slopelaw}, \ref{eq:duration-freq}, and \ref{eq:slope-freq}).

The multi-component bursts were included in Figure \ref{fig:duration-freq} for comparison but were not included in the fits. While we do not expect the same scaling between duration and frequency (and therefore also for the slope and frequency) for these bursts when compared to sub-bursts, they do also follow a clear trend of shorter durations and steeper slopes at higher frequencies. 

As anticipated, the data clearly show unique scaling of these relationships for different sources, as several sources have most of their bursts lying either above or below the trend lines. However, the overall data roughly follow the inverse and quadratic fits, with no sources acting as strong outliers. This suggests that the intrinsic duration $\tau_{\mathrm{w}}'$ and delay time $\tau_{\mathrm{D}}'$, which define these relationships, might be constrained to a given range due to the underlying physical mechanism.

\subsection{Bandwidth vs. Frequency and Bandwidth vs. Duration Relationships}

In the TRDM, FRB emission is intrinsically narrow band at frequency $\nu_0$, with the observed spectral width arising from the large velocity range of the emitting medium \citep{houde2019, Rajabi2020}. If the velocity range is non- or mildly-relativistic, this Doppler broadening yields a linear relationship between the observed bandwidth $\Delta\nu_{\text{obs}}$ and frequency $\nu_{\text{obs}}$ of the form $\Delta\nu_{\text{obs}}\approx 2\Delta\beta'\nu_{\text{obs}}$, where $\Delta\beta'$ represents the rest frame velocity dispersion of the emitting gas\footnote{This linear relationship was derived by performing a second order Taylor expansion for low $\Delta\beta^\prime$ on the relativistic Doppler broadening formula (equation B6 in \citealt{Chamma2021}).}. This is a powerful relationship as it may allow us to probe the dynamics of FRB sources.

Figure \ref{fig:doppler} (top panel) displays the dependence of sub-burst bandwidth $B_\nu$ on central frequency. We fit the data with a linear function $B_\nu=D\nu_{\text{obs}}$, yielding a coefficient of $D=0.138 \pm 0.004$ consistent with $0.14 \pm 0.004$ obtained by \cite{Chamma2022} for FRB 20121102A. An alternative power law fit $D_{\text{free}}\nu_{\text{obs}}^{n_4}$ results in $D_{\text{free}}=0.13 \pm 0.07$ and $n_4=1.01 \pm 0.06$, aligning with the anticipated linear trend.  These data display a clear trend of wider bandwidths at higher frequencies and follow the linear fit closely with no prominent outliers. This is significant as it signifies that all nine FRB sources have similar rest-frame velocity ranges, with $\Delta\beta'\approx 0.07$.

The results of our study confirm both the linear relationship between bandwidth and frequency, and the inverse relationship of frequency with duration as posited by the TRDM. Consequently, we anticipate an inverse correlation between bandwidth and duration, of the form $B_\nu=Et_\mathrm{w}^{-1}$. Based on the previous fits, this coefficient has a predicted value of $E=B\cdot D=206$ MHz ms. This relationship is shown in Figure \ref{fig:doppler} (bottom panel).  However, the visual alignment of our data appears to favor the free power law fit $E_\mathrm{free}t_w^{n_5}$, which yields $E_\mathrm{free}=270\pm20$ and $n_5=-0.53\pm0.05$. Remarkably, these free parameters mirror those identified by \cite{Chamma2022}—a coefficient of $272\pm12$ and an exponent of $-0.53\pm0.04$. This unexpected functionality between the bandwidth and duration is unlikely to be a characteristic of FRB sources but rather the result of propagation effects, and will be explored further in Kumar et al. (in prep). 

\section{Summary}\label{sec:summary}
This study was a comprehensive spectro-temporal analysis of nine repeating FRBs, the largest sample to date. 
\begin{itemize}
    \item We analyzed 175 sub-bursts and 31 multi-component bursts over a frequency range of 149--7144 MHz and a duration range of 73$\mu$s--13.22 ms, using data from seven telescopes. The sources analyzed were FRB 20121102A, 20180916B, 20201124A, 20180301A, 20180814A, 20200929C, 20190804E, 20201130A, and 20190915D.
    \item We tested the predictions of the TRDM, which proposes that FRBs are emitted when a signal from a background source is amplified and triggers strong, narrow band emission in a large region of gas moving relativistically relative to the observer.
    \item We established the sub-burst slope law, an inverse relationship between frequency-normalized sub-burst slope and duration, to hold among sources.
    \item We also discovered the drift rates of multi-component bursts to follow the same trend with duration as sub-bursts, which is not fully explained by the TRDM. It therefore appears that bursts and sub-bursts can be analyzed together, removing the need to split the former into the latter.
    \item We found an inverse relationship between sub-burst duration and frequency, and a quadratic relationship between sub-burst slope and frequency. Significant scatter within each source is consistent with intrinsic spread in the FRB-frame sub-burst duration and delay time, while relative agreement in the fitting coefficients between sources indicates that these timescales may be constrained to a given range.     
    \item We established a linear relationship between sub-burst bandwidth and frequency, consistent with mildly-relativistic Doppler broadening of narrow-band emission with a single rest-frame frequency. Close agreement between the sources suggest consistency among their rest-frame velocity spreads, with $\Delta\beta'\approx0.07$.
    \item We observed an inverse square-root proportionality between sub-burst bandwidth and duration, which is not explained by the TRDM.
\end{itemize}
Overall, the results lend strong support to the TRDM as a viable framework for explaining several aspects of FRB emission. The observed spectro-temporal properties can largely be attributed to the inherent dynamics of the source. Additionally, our findings suggest significant dynamic similarities across sources of repeating FRB, implying shared underlying mechanisms for the phenomena observed.

\section*{Acknowledgements}
M.H.'s research is funded through the Natural Sciences and Engineering Research Council of Canada Discovery Grant RGPIN-2016-04460.

\section*{Data Availability}

The waterfalls of the FRBs can be obtained from the authors of their respective publications. The graphical user interface FRBGUI and related scripts used to perform the measurements are available at \texttt{github.com/mef51/frbgui}.
   

\bibliographystyle{mnras}
\bibliography{main2} 

\begin{thebibliography}{}
\makeatletter
\relax
\def\mn@urlcharsother{\let\do\@makeother \do\$\do\&\do\#\do\^\do\_\do\%\do\~}
\def\mn@doi{\begingroup\mn@urlcharsother \@ifnextchar [ {\mn@doi@}
  {\mn@doi@[]}}
\def\mn@doi@[#1]#2{\def\@tempa{#1}\ifx\@tempa\@empty \href
  {http://dx.doi.org/#2} {doi:#2}\else \href {http://dx.doi.org/#2} {#1}\fi
  \endgroup}
\def\mn@eprint#1#2{\mn@eprint@#1:#2::\@nil}
\def\mn@eprint@arXiv#1{\href {http://arxiv.org/abs/#1} {{\tt arXiv:#1}}}
\def\mn@eprint@dblp#1{\href {http://dblp.uni-trier.de/rec/bibtex/#1.xml}
  {dblp:#1}}
\def\mn@eprint@#1:#2:#3:#4\@nil{\def\@tempa {#1}\def\@tempb {#2}\def\@tempc
  {#3}\ifx \@tempc \@empty \let \@tempc \@tempb \let \@tempb \@tempa \fi \ifx
  \@tempb \@empty \def\@tempb {arXiv}\fi \@ifundefined
  {mn@eprint@\@tempb}{\@tempb:\@tempc}{\expandafter \expandafter \csname
  mn@eprint@\@tempb\endcsname \expandafter{\@tempc}}}

\bibitem[\protect\citeauthoryear{{Bhandari} et~al.,}{{Bhandari}
  et~al.}{2022}]{Bhandari2022}
{Bhandari} S.,  et~al., 2022, \mn@doi [\aj] {10.3847/1538-3881/ac3aec}, \href
  {https://ui.adsabs.harvard.edu/abs/2022AJ....163...69B} {163, 69}

\bibitem[\protect\citeauthoryear{{CHIME/FRB Collaboration} et~al.,}{{CHIME/FRB
  Collaboration} et~al.}{2019}]{Andersen2019}
{CHIME/FRB Collaboration} et~al., 2019, \mn@doi [\apjl]
  {10.3847/2041-8213/ab4a80}, \href
  {https://ui.adsabs.harvard.edu/abs/2019ApJ...885L..24C} {885, L24}

\bibitem[\protect\citeauthoryear{{CHIME/FRB Collaboration} et~al.,}{{CHIME/FRB
  Collaboration} et~al.}{2020}]{Amiri2020}
{CHIME/FRB Collaboration} et~al., 2020, \mn@doi [\nat]
  {10.1038/s41586-020-2398-2}, \href
  {https://ui.adsabs.harvard.edu/abs/2020Natur.582..351C} {582, 351}

\bibitem[\protect\citeauthoryear{{CHIME/FRB Collaboration} et~al.,}{{CHIME/FRB
  Collaboration} et~al.}{2021}]{Amiri2021}
{CHIME/FRB Collaboration} et~al., 2021, \mn@doi [\apjs]
  {10.3847/1538-4365/ac33ab}, \href
  {https://ui.adsabs.harvard.edu/abs/2021ApJS..257...59C} {257, 59}

\bibitem[\protect\citeauthoryear{{CHIME/FRB Collaboration} et~al.,}{{CHIME/FRB
  Collaboration} et~al.}{2023}]{Andersen2023}
{CHIME/FRB Collaboration} et~al., 2023, \mn@doi [\apj]
  {10.3847/1538-4357/acc6c1}, \href
  {https://ui.adsabs.harvard.edu/abs/2023ApJ...947...83C} {947, 83}

\bibitem[\protect\citeauthoryear{{Chamma}, {Rajabi}, {Wyenberg}, {Mathews}  \&
  {Houde}}{{Chamma} et~al.}{2021}]{Chamma2021}
{Chamma} M.~A.,  {Rajabi} F.,  {Wyenberg} C.~M.,  {Mathews} A.,   {Houde} M.,
  2021, \mn@doi [\mnras] {10.1093/mnras/stab2070}, \href
  {https://ui.adsabs.harvard.edu/abs/2021MNRAS.507..246C} {507, 246}

\bibitem[\protect\citeauthoryear{{Chamma}, {Rajabi}, {Kumar}  \&
  {Houde}}{{Chamma} et~al.}{2023}]{Chamma2022}
{Chamma} M.~A.,  {Rajabi} F.,  {Kumar} A.,   {Houde} M.,  2023, \mn@doi
  [\mnras] {10.1093/mnras/stad1108}, \href
  {https://ui.adsabs.harvard.edu/abs/2023MNRAS.522.3036C} {522, 3036}

\bibitem[\protect\citeauthoryear{{Cruces} et~al.,}{{Cruces}
  et~al.}{2021}]{cruces2021}
{Cruces} M.,  et~al., 2021, \mn@doi [\mnras] {10.1093/mnras/staa3223}, \href
  {https://ui.adsabs.harvard.edu/abs/2021MNRAS.500..448C} {500, 448}

\bibitem[\protect\citeauthoryear{{Gajjar} et~al.,}{{Gajjar}
  et~al.}{2018}]{gajjar2018}
{Gajjar} V.,  et~al., 2018, \mn@doi [\apj] {10.3847/1538-4357/aad005}, \href
  {https://ui.adsabs.harvard.edu/abs/2018ApJ...863....2G} {863, 2}

\bibitem[\protect\citeauthoryear{{Gopinath} et~al.,}{{Gopinath}
  et~al.}{2024}]{Gopinath2024}
{Gopinath} A.,  et~al., 2024, \mn@doi [\mnras] {10.1093/mnras/stad3856}, \href
  {https://ui.adsabs.harvard.edu/abs/2024MNRAS.527.9872G} {527, 9872}

\bibitem[\protect\citeauthoryear{{Hessels} et~al.,}{{Hessels}
  et~al.}{2019}]{hessels2019}
{Hessels} J.~W.~T.,  et~al., 2019, \mn@doi [\apjl] {10.3847/2041-8213/ab13ae},
  \href {https://ui.adsabs.harvard.edu/abs/2019ApJ...876L..23H} {876, L23}

\bibitem[\protect\citeauthoryear{{Hewitt} et~al.,}{{Hewitt}
  et~al.}{2022}]{hewitt2021}
{Hewitt} D.~M.,  et~al., 2022, \mn@doi [\mnras] {10.1093/mnras/stac1960}, \href
  {https://ui.adsabs.harvard.edu/abs/2022MNRAS.515.3577H} {515, 3577}

\bibitem[\protect\citeauthoryear{{Hewitt} et~al.,}{{Hewitt}
  et~al.}{2023}]{Hewitt2023}
{Hewitt} D.~M.,  et~al., 2023, \mn@doi [\mnras] {10.1093/mnras/stad2847}, \href
  {https://ui.adsabs.harvard.edu/abs/2023MNRAS.526.2039H} {526, 2039}

\bibitem[\protect\citeauthoryear{{Hilmarsson}, {Spitler}, {Main}  \&
  {Li}}{{Hilmarsson} et~al.}{2021}]{Hilmarsson2021}
{Hilmarsson} G.~H.,  {Spitler} L.~G.,  {Main} R.~A.,   {Li} D.~Z.,  2021,
  \mn@doi [\mnras] {10.1093/mnras/stab2936}, \href
  {https://ui.adsabs.harvard.edu/abs/2021MNRAS.508.5354H} {508, 5354}

\bibitem[\protect\citeauthoryear{{Houde}, {Mathews}  \& {Rajabi}}{{Houde}
  et~al.}{2018}]{Houde2018}
{Houde} M.,  {Mathews} A.,   {Rajabi} F.,  2018, \mn@doi [\mnras]
  {10.1093/mnras/stx3205}, \href
  {https://ui.adsabs.harvard.edu/abs/2018MNRAS.475..514H} {475, 514}

\bibitem[\protect\citeauthoryear{{Houde}, {Rajabi}, {Gaensler}, {Mathews}  \&
  {Tranchant}}{{Houde} et~al.}{2019}]{houde2019}
{Houde} M.,  {Rajabi} F.,  {Gaensler} B.~M.,  {Mathews} A.,   {Tranchant} V.,
  2019, \mn@doi [\mnras] {10.1093/mnras/sty3046}, \href
  {https://ui.adsabs.harvard.edu/abs/2019MNRAS.482.5492H} {482, 5492}

\bibitem[\protect\citeauthoryear{{Jahns} et~al.,}{{Jahns}
  et~al.}{2023}]{Jahns2022}
{Jahns} J.~N.,  et~al., 2023, \mn@doi [\mnras] {10.1093/mnras/stac3446}, \href
  {https://ui.adsabs.harvard.edu/abs/2023MNRAS.519..666J} {519, 666}

\bibitem[\protect\citeauthoryear{{Josephy} et~al.,}{{Josephy}
  et~al.}{2019}]{josephy2019}
{Josephy} A.,  et~al., 2019, \mn@doi [\apjl] {10.3847/2041-8213/ab2c00}, \href
  {https://ui.adsabs.harvard.edu/abs/2019ApJ...882L..18J} {882, L18}

\bibitem[\protect\citeauthoryear{{Lorimer}, {Bailes}, {McLaughlin}, {Narkevic}
  \& {Crawford}}{{Lorimer} et~al.}{2007}]{Lorimer2007}
{Lorimer} D.~R.,  {Bailes} M.,  {McLaughlin} M.~A.,  {Narkevic} D.~J.,
  {Crawford} F.,  2007, \mn@doi [Science] {10.1126/science.1147532}, \href
  {https://ui.adsabs.harvard.edu/abs/2007Sci...318..777L} {318, 777}

\bibitem[\protect\citeauthoryear{{Luo} et~al.,}{{Luo} et~al.}{2020}]{luo2020}
{Luo} R.,  et~al., 2020, \mn@doi [\nat] {10.1038/s41586-020-2827-2}, \href
  {https://ui.adsabs.harvard.edu/abs/2020Natur.586..693L} {586, 693}

\bibitem[\protect\citeauthoryear{Lyutikov, Burzawa  \& Popov}{Lyutikov
  et~al.}{2016}]{lyutikov2016}
Lyutikov M.,  Burzawa L.,   Popov S.~B.,  2016, \mn@doi [\mnras]
  {10.1093/mnras/stw1669}, 462, 941

\bibitem[\protect\citeauthoryear{{Marcote} et~al.,}{{Marcote}
  et~al.}{2020}]{marcote2020}
{Marcote} B.,  et~al., 2020, \mn@doi [\nat] {10.1038/s41586-019-1866-z}, \href
  {https://ui.adsabs.harvard.edu/abs/2020Natur.577..190M} {577, 190}

\bibitem[\protect\citeauthoryear{{Marthi}, {Gautam}, {Li}, {Lin}, {Main},
  {Naidu}, {Pen}  \& {Wharton}}{{Marthi} et~al.}{2020}]{marthi2020}
{Marthi} V.~R.,  {Gautam} T.,  {Li} D.~Z.,  {Lin} H.~H.,  {Main} R.~A.,
  {Naidu} A.,  {Pen} U.~L.,   {Wharton} R.~S.,  2020, \mn@doi [\mnras]
  {10.1093/mnrasl/slaa148}, \href
  {https://ui.adsabs.harvard.edu/abs/2020MNRAS.499L..16M} {499, L16}

\bibitem[\protect\citeauthoryear{{Michilli} et~al.,}{{Michilli}
  et~al.}{2018}]{michilli2018}
{Michilli} D.,  et~al., 2018, \mn@doi [\nat] {10.1038/nature25149}, \href
  {https://ui.adsabs.harvard.edu/abs/2018Natur.553..182M} {553, 182}

\bibitem[\protect\citeauthoryear{{Nimmo} et~al.,}{{Nimmo}
  et~al.}{2021}]{Nimmo2021}
{Nimmo} K.,  et~al., 2021, \mn@doi [Nature Astronomy]
  {10.1038/s41550-021-01321-3}, \href
  {https://ui.adsabs.harvard.edu/abs/2021NatAs...5..594N} {5, 594}

\bibitem[\protect\citeauthoryear{{Nimmo} et~al.,}{{Nimmo}
  et~al.}{2022}]{Nimmo2022}
{Nimmo} K.,  et~al., 2022, \mn@doi [\apjl] {10.3847/2041-8213/ac540f}, \href
  {https://ui.adsabs.harvard.edu/abs/2022ApJ...927L...3N} {927, L3}

\bibitem[\protect\citeauthoryear{{Petroff}, {Hessels}  \& {Lorimer}}{{Petroff}
  et~al.}{2022}]{Petroff2022}
{Petroff} E.,  {Hessels} J.~W.~T.,   {Lorimer} D.~R.,  2022, \mn@doi [\aapr]
  {10.1007/s00159-022-00139-w}, \href
  {https://ui.adsabs.harvard.edu/abs/2022A&ARv..30....2P} {30, 2}

\bibitem[\protect\citeauthoryear{{Pleunis} et~al.,}{{Pleunis}
  et~al.}{2021}]{pleunis2021}
{Pleunis} Z.,  et~al., 2021, \mn@doi [\apjl] {10.3847/2041-8213/abec72}, \href
  {https://ui.adsabs.harvard.edu/abs/2021ApJ...911L...3P} {911, L3}

\bibitem[\protect\citeauthoryear{{Price} et~al.,}{{Price}
  et~al.}{2019}]{Price2019}
{Price} D.~C.,  et~al., 2019, \mn@doi [\mnras] {10.1093/mnras/stz958}, \href
  {https://ui.adsabs.harvard.edu/abs/2019MNRAS.486.3636P} {486, 3636}

\bibitem[\protect\citeauthoryear{{Rajabi}, {Chamma}, {Wyenberg}, {Mathews}  \&
  {Houde}}{{Rajabi} et~al.}{2020}]{Rajabi2020}
{Rajabi} F.,  {Chamma} M.~A.,  {Wyenberg} C.~M.,  {Mathews} A.,   {Houde} M.,
  2020, \mn@doi [\mnras] {10.1093/mnras/staa2723}, \href
  {https://ui.adsabs.harvard.edu/abs/2020MNRAS.498.4936R} {498, 4936}

\bibitem[\protect\citeauthoryear{{Rajwade} et~al.,}{{Rajwade}
  et~al.}{2020}]{Rajwade2020}
{Rajwade} K.~M.,  et~al., 2020, \mn@doi [\mnras] {10.1093/mnras/staa1237},
  \href {https://ui.adsabs.harvard.edu/abs/2020MNRAS.495.3551R} {495, 3551}

\bibitem[\protect\citeauthoryear{{Tendulkar} et~al.,}{{Tendulkar}
  et~al.}{2017}]{Tendulkar2017}
{Tendulkar} S.~P.,  et~al., 2017, \mn@doi [\apjl] {10.3847/2041-8213/834/2/L7},
  \href {https://ui.adsabs.harvard.edu/abs/2017ApJ...834L...7T} {834, L7}

\bibitem[\protect\citeauthoryear{{Wang}, {Yang}, {Niu}, {Xu}  \&
  {Zhang}}{{Wang} et~al.}{2022}]{wang2022_freq}
{Wang} W.-Y.,  {Yang} Y.-P.,  {Niu} C.-H.,  {Xu} R.,   {Zhang} B.,  2022,
  \mn@doi [\apj] {10.3847/1538-4357/ac4097}, \href
  {https://ui.adsabs.harvard.edu/abs/2022ApJ...927..105W} {927, 105}

\makeatother
\end{thebibliography}

\bsp	
\label{lastpage}
\end{document}